\documentclass[useAMS,usenatbib]{mn2e}
\usepackage{psfig, epsf, epsfig}
\title[Collision with the invisible]{Formation of the off-center
bar in the Large Magellanic Cloud: A collision with a dark satellite ?} 
\author[K. Bekki ]{Kenji Bekki${}^1$\thanks{E-mail:
bekki@phys.unsw.edu.au} \\
       ${}^1$School of Physics, University of New South Wales,
              Sydney 2052, NSW, Australia\\}

\begin{document}

\date{Accepted, Received 2005 February 20; in original form }

\pagerange{\pageref{firstpage}--\pageref{lastpage}} \pubyear{2005}

\maketitle

\label{firstpage}

\begin{abstract}

Recent observations on structural 
properties of the  Large Magellanic
Cloud (LMC) based on
the Deep Near-Infrared Southern Sky Survey (DENIS)
and Two Micron All-Sky Survey (2MASS) have revealed
that the LMC has an  off-center  bar even in
the older stellar populations.
Previous dynamical models including tidal
interaction between the LMC, the Small Magellanic Cloud (SMC),
and the Galaxy, however,   did not  reproduce
so well the older off-center bar.
We here show that the off-center bar can be formed
if the LMC with an already existing bar can 
collide  with a low-mass Galactic 
subhalo as massive  as $\sim 5 \times 10^8 {\rm M}_{\odot}$ 
(corresponding roughly to a few \% of the LMC mass).
The simulated  stellar distribution after the 
collision appears to show an ``off-center bar'',
not because the center of the bar significantly
deviates from the dynamical center of the LMC,
but because the underlying stellar
distribution of the disk is significantly asymmetric
with respect to the center of the bar. 
We discuss whether off-center bars observed in Magellanic-type
dwarf galaxies can be formed as a result of tidal interaction
with low-mass halos with no or little visible matters.

\end{abstract}

\begin{keywords}
Magellanic Clouds -- galaxies:structure --
galaxies:kinematics and dynamics -- galaxies:halos -- galaxies:star
clusters
\end{keywords}

\section{Introduction}

Recent photometric and spectroscopic observations on structures
and kinematics of stars with different ages
and metallicities in the LMC  have significantly  improved our
understanding of the dynamical properties
(e.g., Cioni et al. 2000; van der Marel 2001, v01;
Minniti et al. 2003; Alves 2004; 
Cole et al. 2005;
Grocholski et al. 2006; 
Olsen \& Massey 2007).
For example,
observational studies based on
the DENIS
and the 2MASS have revealed
that the LMC has the off-center stellar bar in a significantly
elongated stellar disk (e.g., v01).
Although the spatial distribution of these intermediate-age
stellar populations (e.g., AGB/RGB stars) does not show clearly
spiral arms (v01), younger stellar populations are observed
to show peculiar arms which may have formed from the past
tidal interaction between the LMC and the SMC (Olsen \& Massey 2007).
These observations imply that stellar populations
with different ages have different spatial distributions
in the LMC.

\begin{table*}
\centering
\begin{minipage}{185mm}
\caption{Model parameters for DMSH-LMC interaction.} 
\begin{tabular}{cccccc}
Model name & 
{$M_{\rm dmsh}$ ($\times M_{\rm t}$)
\footnote{ The mass of a dark matter subhalo (DMSH) in units of
the mass of the LMC ($M_{\rm t}$).}}  & 
{$r_{\rm p}$ ($\times R_{\rm d}$)
\footnote{ The pericenter distance of the DMSH  in units of
the LMC disk size ($R_{\rm d}$).}}  & 
Orbital eccentricity ($e$)  &
Inclination angle (${\theta}^{\circ}$) &
Comments \\
M1 & 0.05 & 0.5 & 1.0 & 180 & the standard model \\
M2 & 0.01 & 0.5 & 1.0 & 180 & less massive  DMSH \\
M3 & 0.02 & 0.5 & 1.0 & 180 &  \\
M4 & 0.1 & 0.5 & 1.0 & 180 & more massive DMSH \\
M5 & 0.05 & 0.5 & 1.0 & 0 & prograde encounter \\
M6 & 0.05 & 1.0 & 1.0 & 180 & larger $r_{\rm p}$   \\
M7 & 1.0 & 3.0 & 0.5 & 180 & distant encounter  \\
M8 & 0.05 & 0.01 & 1.0 & 90 & almost head-on collision  \\
M9 & 0.1 & 0.01 & 1.0 & 90 &   \\
M10 & - & - & - & - & isolated model
\end{tabular}
\end{minipage}
\end{table*}

One of intriguing results  as to the dynamical properties
of the LMC in these observations
is the off-center bar seen in the projected distribution
of {\it intermediate-age  stellar populations}: 
a prominent off-center bar embedded within its flat disk component
was already identified early 
for the optical image determined largely by  the distributions
of younger stellar populations  
(e.g., de Vaucouleurs \& Freeman 1972). 
Recent numerical simulations have shown that the
two-dimensional distribution of
the $B-$band surface brightness (${\mu}_{\rm B}$)
in the LMC  stellar disk after the LMC-SMC-Galaxy interaction
about 0.2 Gyr ago appears to  have an off-center bar 
to some extent (Bekki \& Chiba 2007; BC07).
However, 
the simulated  asymmetric 2D distribution with apparently an off-center
bar 
is  due largely to young stars formed in the asymmetric gas (BC07):
the spatial distribution
of the older stellar populations is highly unlikely to show an off-center 
bar.
Other numerical simulations on dynamical evolution of the LMC
also failed to reproduce well the off-center bar 
(e.g., Bekki \& Chiba 2005, BC05; Mastropietro et al. 2005).
Thus the origin of the off-center bar is far from being understood well
in the context of the past  dynamical interaction between the 
Magellanic Clouds (MCs)
and the Galaxy.

The purpose of this Letter is to show,
for the first time,  that if
the LMC can collide with a low-mass subhalo 
of the Galaxy, 
the LMC can develop an off-center bar in the old disk.
We discuss the masses and orbits of the subhalos required for
the formation of the off-center bar based on
a large parameter survey of tidal interaction
between the LMC and the dark matter subhalos (DMSHs).
The time scale of a LMC-DMSH merger/collision event ($t_{\rm m}$)
can be estimated
as follows (e.g., Makino \& Hut 1997);
\begin{equation}
t_{\rm m}=\frac{ 1 } {n_{\rm h}\sigma v},
\end{equation}
where $n_{\rm h}$, $\sigma$, and $v$
are the mean number density of the DMSHs within 
the Galaxy,
the geometrical cross section of the LMC ,
and a relative velocity between a DMSH  and the LMC.
We here estimate $n_{\rm h}$ for the central 50 kpc
of the Galaxy (corresponding to the pericenter of the LMC orbit)
and assume that  $\sigma= \pi {R_{\rm d}}^2$,
where $R_{\rm d}$ is the LMC size
and $v$ is velocity dispersion ($=v_{\rm c}/\sqrt(2)$,
where $v_{\rm c}$ is the circular velocity thus 220 km s$^{-1}$)  of the Galaxy
halo.
If we use the results of
the latest cosmological simulations
(e.g., Springel et al. 2008), which can resolve 300,000 subhalos
of the Galaxy, 
we can estimate that the total number of subhalos ($N_{\rm h}$)
with masses larger than $\sim 5 \times 10^8 {\rm M}_{\odot}$
within $50-100$ kpc is about 20 (from the results in their
Figs. 9 and 11). 
Then we can derive $t_{\rm m}$ as follows; 
\begin{equation}
t_{\rm m}= 2.1 
{ ( \frac{ N_{\rm h} } {20} )  }^{-1}
{ ( \frac{ R_{\rm d} } {5 {\rm kpc}} )  }^{-2}
{ ( \frac{ v } {156  {\rm km s^{-1} } } )  }^{-1}
{\rm Gyr}
\end{equation}
This suggests that the LMC-DMSH collision is not so rare
and thus worth a numerical investigation.

\section{The model}

We numerically investigate dynamical impact of a 
DMSH in the Galaxy
on the evolution of the stellar disk of in the LMC by using our original
GRAPE-SPH code (e.g., Bekki \& Chiba 2006). Since we focus exclusively
on the tidal effect of the subhalo in the present study, 
we investigate dynamical evolution of {\it a purely collisionless system}
for  variously different model parameters of the DMSH-LMC interaction. 
In order to show more clearly the dynamical impact of the DMSH-LMC interaction,
we do not include any other tidal effects of the Galaxy and the SMC:
non-inclusion of these effects greatly helps us to grasp essential ingredients
of the tidal effect from a DMSH. 

Since the details of the models for the LMC
and numerical methods and techniques  are given
in BC05, we here briefly describe them.
The LMC is modeled as a fully self-gravitating system
and  composed of a live dark halo
and a thin exponential disk with no bulge.
The total mass of the dark halo,
that of the disk,  and the size of the disk  
are $M_{\rm dm}$, $M_{\rm d}$,
and $R_{\rm d}$, respectively. 
The mass ratio of the dark halo 
to the total mass is
fixed at 0.7 throughout the paper, which is consistent with
the observation by v01.
We show the results of the models with the total mass 
($M_{\rm t} =M_{\rm dm}$ + $M_{\rm d}$)
of the LMC being $2\times 10^{10} {\rm M}_{\odot}$ 
within the tidal radius ($r_{\rm t}$) of 15 kpc (e.g., v01, BC05)
in the present study.

We use the ``NFW profile'' (Navarro, Frenk \& White 1996)
adopted in B05 for the radial density profile of the dark matter
halo of the LMC and distribute the dark matter particles 
only within $r_{\rm t}$.
The radial ($R$) and vertical ($Z$) density profile 
of the initially thin  disk of the LMC are  assumed to be
proportional to $\exp (-R/R_{0}) $ with scale length $R_{0}$ 
= $0.2 R_{\rm d}$
and to  ${\rm sech}^2 (Z/Z_{0})$ with scale length $Z_{0}$ = $0.2R_{0}$, 
respectively.
In addition to the rotational velocity made by the gravitational
field of disk and halo component, the initial radial 
and azimuthal velocity
dispersion are given to the disk component according
to the epicyclic theory with Toomre's parameter 
(Binney \& Tremaine 1987) $Q$ = 1.5.

The simulations have  mass and size  resolutions  of  
$2 \times 10^6 {\rm M}_{\odot}$ and $105$ pc for stars
for models with particle numbers of a quarter million
and $R_{\rm d}=5$ kpc. 
The gravitational softening length (${\epsilon}_{\rm dmsh}$)
of a subhalo is assumed to
be different from those of dark matter and stellar particles,
and models with different ${\epsilon}_{\rm dmsh}$ 
are investigated.
We confirm that the present results do not depend
on the adopted Plummer softening lengths ranging from 140pc
to 500pc.

A DMSH is modeled as a point-mass particle with 
a mass ($M_{\rm dmsh}$)
ranging from $0.01 M_{\rm t}$ to $1 M_{\rm t}$
and assumed to have no baryonic components such
as cold gas and stars. The pericenter distances ($r_{\rm p}$)  of
the DMSH-LMC interaction is a free parameter ranging
from $0.01R_{\rm d}$ to $3R_{\rm d}$ in the present study.
The initial spin of a LMC disk in  a model is specified by two angles,
$\theta$ and $\phi$, where
$\theta$ is the angle between the $Z$-axis and the vector of
the angular momentum of a disk and
$\phi$ is the azimuthal angle measured from $X$-axis to
the projection of the angular momentum vector
of a disk onto the $X-Y$ plane.
Prograde and retrograde orbits have $\theta = 0^{\circ}$
and  $\theta = 180^{\circ}$, respectively, for $\phi=0^{\circ}$
in the above definition.
In order to show more clearly the importance of orbital spins
in the DMSH-LMC interaction,
we present the results of the models with $\theta = 0^{\circ}$
or $\theta = 180^{\circ}$ and $\phi=0^{\circ}$.

We consider that the lopesided distribution of older stars in
the LMC (v01) is short-lived and thus can be formed from
recent tidal interaction with a DMSH within less than 1 Gyr ago. 
Since the observed bar contains a significant fraction
of intermediate-age stars (v01;  Smecker-Hane et al. 2002),
we need to consider that
the LMC before the tidal interaction 
has a well developed stellar bar.
Therefore we  first dynamically relax the LMC disk for $\sim 7$ Gyr
and then we use the final barred stellar distribution
(formed during the first relaxation process)  as the initial
one for the tidal interaction between a barred disk and a DMSH.
This dynamically relaxed disk has a higher velocity
dispersion and thus no spiral arms owing  to
long-term dynamical heating by the developed bar so that the disk
can be more consistent with the observed one with no spirals
for older AGB and RGB stars (v01).

Although we have investigated a large number (30) of models
with different $M_{\rm dmsh}$, $r_{\rm p}$, and $\theta$,
we mainly show the results of ``the standard model'' in which
$M_{\rm dmsh}=0.05 M_{\rm t}$,
$r_{\rm p}=0.5 R_{\rm d}$, and 
$\theta = 180^{\circ}$.
This is because this model can show more clearly a typical behavior
of the  off-center bar formation.
In order to explain briefly dependences of the results on model
parameters, we show the results of the selected ten  models,
for which the parameter values are shown in the Table 1.
Although it is pretty obvious that the projected
stellar distribution in the LMC
shows an off-center bar (e.g., de Vaucouleurs \& Freeman 1972; v01),
previous observations did not quantify the degree of
``off-centeredness''  for the bar.
We thus discuss the morphologies
of the simulated  two-dimensional (2D) distribution
of older stars
in a more qualitative manner in the present paper: quantitative discussion
on how to define the off-centeredness of a bar needs to be done
in a future paper.
The surface mass-densities of old stars
(${\mu}_{\rm s}$) is defined as 
 ${\log}_{10} 
{\Sigma}_{\rm s}$, where ${\Sigma}_{\rm s}$ is the
projected mass density (in units of ${\rm M}_{\odot}$ pc$^{-2}$)
at each  of  $50 \times 50$ cells for a region
of 11 kpc $\times$ 11 kpc in a model.
We derive the smoothed 2D fields of ${\mu}_{\rm s}$  
by using the same smoothing method adopted in our previous
paper (Bekki \& Peng 2006).

\begin{figure}
\psfig{file=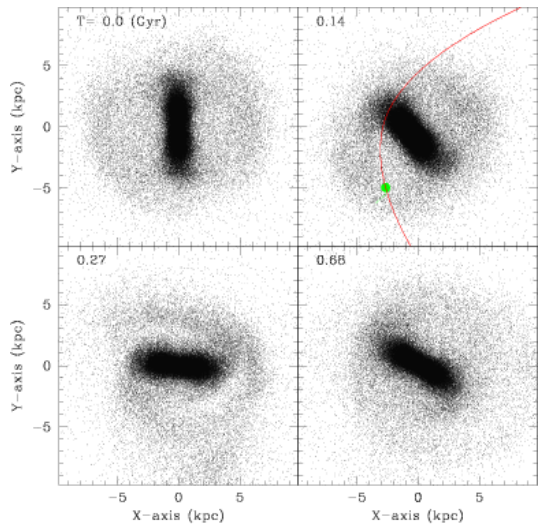,width=7.cm}
\caption{
Time evolution of stellar distributions in the LMC projected onto
the $x$-$y$ plane for the standard model M1.
The dynamical center of the LMC is coincident with the center 
of each frame and the time $T$ in units of  Gyr is shown
in the upper left corner of each frame. The DMSH and its
orbit are shown by a big green circle and a red solid line,
respectively, in the upper right  panel for $T=0.14$ Gyr
when the DMSH collides with the LMC.
Note that this retrograde tidal perturbation during the collision
can cause an asymmetric distribution without forming
strong spiral arms.
}  
\label{Figure. 1}
\end{figure}

\begin{figure}
\psfig{file=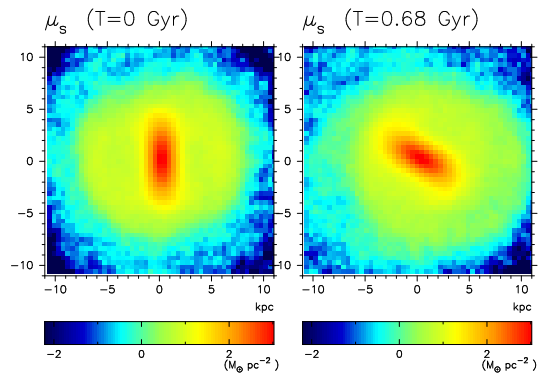,width=7.cm}
\caption{
Initial (left) and final (right) 2D distributions 
of ${\mu}_{\rm s}$ in units of ${\rm M}_{\odot}$ pc$^{-2}$
(logarithmic scale)
for the standard model. 
For convenience, the dynamical center
(i.e., the center of the bar)
is set to coincide with the center of each frame.
It appears that
the ${\mu}_{\rm s}$ distribution after the DMSH-LMC collision
is lopesided with respect to the center of the bar.
This means that the bar appears to be off-center with respect
to the center of the disk.
}  
\label{Figure. 2}
\end{figure}

\begin{figure}
\psfig{file=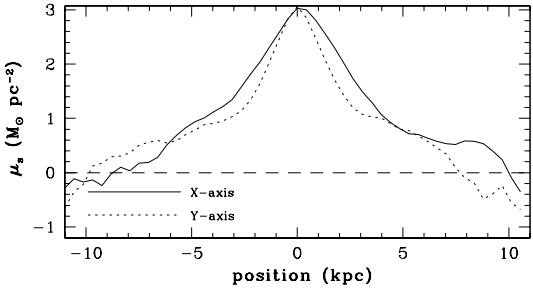,width=7.cm}
\caption{
The radial profiles 
of ${\mu}_{\rm s}$ 
along $x$-axis (solid, for particles with $|y| \le 220$pc) 
and $y$-axis (dotted, for particles with $|x| \le 220$pc)
for the stellar disk at $T=0.68$ Gyr in the standard model.
For comparison,  
(${\mu}_{\rm s} = {\log}_{10} {\Sigma}_{\rm s}
\sim 0  {\rm M}_{\odot}$ pc$^{-2}$)
is shown by a dashed line.
Note that the outer disk ($|x|>5$ kpc or $|y|>5$ kpc)
shows asymmetric distributions.
}  
\label{Figure. 3}
\end{figure}

\begin{figure}
\psfig{file=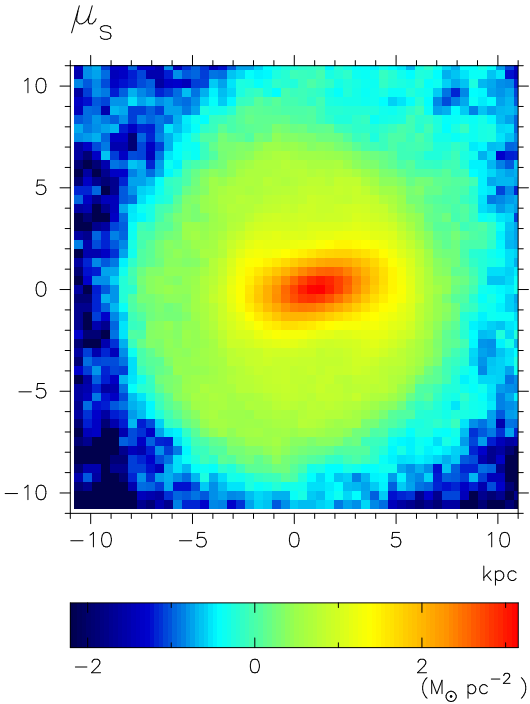,width=7.cm}
\caption{
The 2D distribution 
of  ${\mu}_{\rm s}$  
at $T=0.68$ Gyr for the stellar disk that is rotated and inclined
such that the distribution can be similar to the observed distribution
of AGB/RGB stars in the LMC
by v01. Here the center of the frame is set to be coincident with
the center of the iso-density for the outer part of the simulated disk
(${\mu}_{\rm s} = {\log}_{10} {\Sigma}_{\rm s}
\sim 0.5  {\rm M}_{\odot}$ pc$^{-2}$): the center of the bar
deviates  $\sim$1 kpc from the isophotal center. 
The disk appears to have an off-center bar in this 2D distribution.
}  
\label{Figure. 4}
\end{figure}

\begin{figure}
\psfig{file=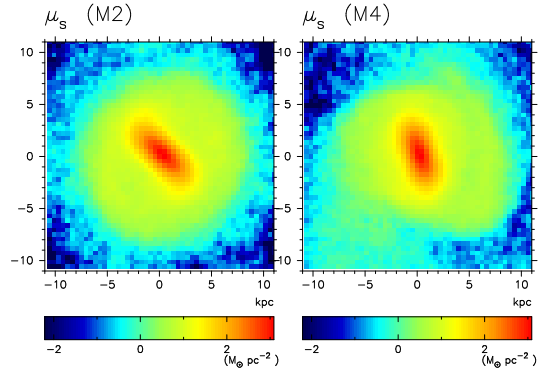,width=7.cm}
\caption{
The same as Fig. 2 but for the models M2 (left)
with $M_{\rm dmsh}=0.01 M_{\rm t}$ 
and M4 (right) with  $M_{\rm dmsh}=0.1 M_{\rm t}$.
}  
\label{Figure. 5}
\end{figure}

\section{Results}

Fig. 1 shows how the moderately strong dynamical impact during
the collision between a DMSH and the LMC can change the overall
distribution of stars in the LMC for the standard model M1.
As shown in Fig. 1,
the tidal perturbation of the  DMSH-LMC collision can change
the stellar distribution in the outer part of the LMC disk,
though it can not change the shape of the central stellar bar.
Owing to the retrograde tidal encounter,
strong bisymmetric spiral arms can not be formed during the collision
in this model. The center of the bar is coincident with the 
dynamical center of the LMC (i.e., the center of the mass)
during and after the DMSH-LMC collision. The stellar distribution
in the outer part of 
the disk, on the other hand, appears to be lopesided with
respect to the center of the bar at
$T=0.27$ and 0.68 Gyr,
which means that the simulated LMC after the collision
appears to have an off-center bar.
The stellar vertical structure and kinematics do not change during the collision.

Fig. 2 shows the 2D distributions of ${\mu}_{\rm s}$ 
at $T=0$ and 0.68 Gyr in the standard model.
It is clearer  in Fig. 2 than in Fig. 1 that  
the final stellar distribution of the disk at $T=0.68$ Gyr
appears to have an off-center bar: the center of the iso-density
contour of ${\mu}_{\rm s} \sim 0.5 {\rm M}_{\odot} {\rm pc}^{-2}$
is  not coincident with the center of the bar.
Fig. 3 confirms that the ${\mu}_{\rm s}$ distributions
along the $x$- and $y$-axes 
around $R=5-10$ kpc
are appreciably asymmetric.
These results mean that the disk appears to have an off-center bar,
not because the center of the bar really deviates from 
the center of the mass for the disk,
but because 
the stellar distribution in the outer part of the disk
is lopesided with respect to the center of the bar.
These results imply that the origin 
of the off-center bar in the LMC is due largely
to the lopesided distribution of stars
in the outer part of the LMC {\it with respect to the center of the bar}:the
dynamical center of the LMC is coincident with the center of the 
apparently off-center bar.

It is found that 
the simulated 2D distribution of ${\mu}_{\rm s}$ 
at each time $T$ depends strongly
on viewing angles of the LMC disk.
Fig. 4 shows the final 2D distribution of ${\mu}_{\rm s}$ 
(i.e., $T=0.68$ Gyr in M1)
that is rotated by 220$^{\circ}$ in the $x$-$y$ plane
and then inclined by 30$^{\circ}$ with respect to the $x$-$y$ plane.
These rotation and inclination angles are chosen such that
the distribution can be compared with  the observed one for older stellar
populations by v01.
It is clear in Fig. 4 that the center of
the bar appears to deviate from the  center of the disk
(which is defined by the center of the iso-density contour
for  ${\mu}_{\rm s} \sim 0.5 {\rm M}_{\odot} {\rm pc}^{-2}$).
Thus these results suggest that the origin of the off-center bar
in the LMC can be closely associated with the past collision
between a DMSH and the LMC.

Parameter dependences of the present study are summarized as follows.
Firstly, only the models with $M_{\rm dmsh} \ge 0.02 M_{\rm t}$ 
can show apparently off-center bars after the DMSH-LMC collision.
Fig. 5 shows that the final disk of the LMC
in the model with  $M_{\rm dmsh} = 0.01 M_{\rm t}$ (M2)
does not have an off-center bar whereas the disk 
in the model with $M_{\rm dmsh} = 0.1 M_{\rm t}$ (M4)
has an apparently off-center bar and a significantly 
asymmetric distribution in its outer part.
The absence of a remarkable off-center bar in the model M2
suggests that there is a threshold  $M_{\rm dmsh}$ for
the off-center bar formation by the DMSH-LMC collision.
The final ${\mu}_{\rm s}$ distribution in the model M4
is too asymmetric to be consistent with the observed one by v01,
which suggests that there can be a range of  $M_{\rm dmsh}$
required for explaining well the observed stellar distribution
with a less disturbed morphology in the LMC.

Secondly, the models with retrograde encounters (e.g., M1)
can show more clearly  off-center bars
than those with prograde encounters (e.g., M5). 
Thirdly,  the models with larger  $r_{\rm p}$ ($ \ge R_{\rm d}$) do
not show off-center bars (M6), which means that there is a threshold
$r_{\rm p}$ below which an off-center bar can be formed from
the DMSH-LMC collision. 
Fourthly, the models with large $M_{\rm dmsh}$ and large $r_{\rm p}$
(i.e., distant encounters with more massive subhalos) do not
show remarkable off-center bars (M7).
The model M8 in which a DMSH can make an almost head-on collision
from the polar-axis of the LMC shows an off-center bar 
more clearly {\it just after}
the collision without forming rings and spirals: the off-center
appearance is short-lived. 
This result implies that if the off-center bar of the LMC
can form from an almost head-on collision between a DMSH
and the LMC,  the collision  should happen very recently (i.e.,
within less than 0.1 Gyr).
Formation processes of the off-center bar in M8 are 
essentially the same as those of off-center bars/nuclei
seen in colliding ring galaxies like NGC 922 (e.g., Wong et al. 2006)

\section{Discussion and conclusions}

We consider that the LMC-SMC-Galaxy
interaction alone can not explain the off-center bar seen
in {\it older stellar populations} (v01) in the present study.
It should be, however, stressed that previous simulations
(e.g., BC05; BC07; Mastropietro et al. 2005)
that failed to reproduced the off-center bar  in 
{\it older stars} of the LMC
have not explored past orbits of the MCs consistent with
the latest proper motion measurements (e.g., K06):
there could be some orbital models consistent with
K06 in which 
the SMC can interact/collide  with the LMC in a retrograde
sence  and therefore induce the off-center bar formation 
LMC without forming remarkable spirals. 
Accordingly, although previous simulations 
did not support  retrograde tidal interaction between
the MCs  (BC07),
it would be fair to say currently that tidal perturbation from the SMC
can not be completely ruled out as the mechanism for the off-center bar
formation.

We have first shown that the observed peculiar non-axisymmetric
structure (i.e., off-center bar with no remarkable spirals)
in the older stellar populations
can  be due to the past interaction
between the LMC and a low-mass subhalo with the mass  
as large as $\sim 5$\% of the total mass of the LMC
(i.e., $M_{\rm dmsh} \sim 5 -10 \times 10^{8} {\rm M}_{\odot}$).
Recent observational and theoretical studies on the
ultra-faint dwarf galaxies have suggested that
there can be a minimum halo mass of $10^9 {\rm M}_{\odot}$
for the formation of galaxies (Strigari et al. 2008):  
dark halos with masses less
than $10^9 {\rm M}_{\odot}$ can have little or no visible
matter.  Thus the present study suggests that
the low-mass subhalo possibly  responsible for
the formation of the off-center bar in the LMC
would have no or little
visible matter (such subhalos are referred to as ``dark satellites''
for convenience  from now on).

Previous numerical simulations showed that the dynamical impact
of a dark satellite on the HI disk of a galaxy interacting with
the satellite (``dark impact'')
can be responsible not only for the formation of
a giant HI hole  and filamentary structures in the HI disk
but also for star formation in the disk
(Bekki \& Chiba 2006). 
It would be therefore possible that one of the observed
HI holes (Stavely-Smith et al. 2003), the filamentary
structures connecting  to the 30 Doradus region
(Nidever et al. 2008),
and LMC 4 (e.g., Efremov 2004)  were formed as a result of
the collision between the LMC and a dark satellite.
Since the dark impact on the LMC may well
change the orbit, the orbital evolution of the LMC
for the Magellanic stream model
would need to be reconsidered in the context of
the possible LMC-DMSH interaction.

Although dark satellites with masses of $10^8-10^9 {\rm M}_{\odot}$
can not strongly disturb the disks of luminous
galaxies like the Galaxy,
they can significantly influence less massive galaxies like the LMC.
Although barred Magellanic-type dwarfs are not
rare (e.g., Freeman 1984),
the origin of off-center bars in these dwarfs,
remains unclear (e.g., Wilcots \& Prescott 2004).
The present results strongly suggest that
even apparently isolated Magellanic dwarfs with off-center bars
might have been formed from past interaction between the dwarfs
and their dark satellites.
The observed very peculiar morphology of the LMC appears to have
made us realize a possibly important effect of dark satellites
on evolution of less massive galaxies.

\section{Acknowledgment}
I am   grateful to the anonymous referee for valuable comments,
which contribute to improve the present paper.
K.B. acknowledges the Large Australian Research Council (ARC).
Numerical computations reported here were carried out on the GRAPE system
at the University of New South Wales  and that 
kindly made available by the Center for computational
astrophysics
(CfCA) of the National Astronomical Observatory of Japan.

\end{document}